# A Comparison Between Decision Trees and Decision Tree Forest Models for Software Development Effort Estimation

[1]Ali Bou Nassif and [2]Mohammad Azzeh

[1]University of Western Ontario, London, ON, Canada
[1]abounas@uwo.ca
[2]Applied Science University, Amman, Jordan
[2]m.y.azzeh@asu.edu.jo

[3]Luiz Fernando Capretz and [4]Danny Ho

[3]University of Western Ontario, London, ON, Canada
[3]lcapretz@uwo.ca
[4]NFA Estimation Inc., Richmond Hill, Canada
[4]danny@nfa-estimation.com

*Abstract*—Accurate software effort estimation has been a challenge for many software practitioners and project managers. Underestimation leads to disruption in the project's estimated cost and delivery. On the other hand, overestimation causes outbidding and financial losses in business. Many software estimation models exist; however, none have been proven to be the best in all situations. In this paper, a decision tree forest (DTF) model is compared to a traditional decision tree (DT) model, as well as a multiple linear regression model (MLR). The evaluation was conducted using ISBSG and Desharnais industrial datasets. Results show that the DTF model is competitive and can be used as an alternative in software effort prediction.

## I. INTRODUCTION

Software effort estimation is crucial at the inception of each software project. This will help project managers effectively bid on projects and supervise winning projects. Software estimators have been notorious in predicting unrealistic software effort and cost [1]. The Standish Chaos Report [2] states that about 65% of software projects are not delivered on time and within budget. The main attribute /factor that correlates to software effort is software size; however, other factors are also important such as the degree of fitness between the person and the organization [3], [4]. Reusability also plays an important role since software projects with reused components require lesser effort to be developed [5].

Many models have been used in the last five decades to predict software effort. These include algorithmic models such as COCOMO [6], SLIM [7] and regression [8], [9], expert judgment such as [10], estimation using analogy such as [11] and [12], soft computing and machine learning models such as fuzzy logic, neural network, genetic algorithms. Examples of the latter include [13], [14], [15], [16], [17], [18] and [19]. In spite of the numerous software effort prediction models, each one has its advantages and disadvantages. Research has shown that some software estimation models excel in some areas and based on certain datasets, but they fail to provide good accuracy in different problem domains.

Decision tree (DT) models, also known as classification and regression trees (CART) have been widely used in data mining and machine learning due to its simplicity, as one need not be an expert to comprehend and interpret the tree.

Experts have developed models which are based on CART, such as fuzzy regression trees, Treeboost, model trees and forest trees to improve the accuracy of the traditional CART models. In this paper, we compare a decision tree forest (DTF) model to a traditional decision tree (DT) model, as well as a multiple linear regression (MLR) model. The comparison was performed using ISBSG and Desharnais industrial datasets and based on common software effort evaluation criteria such as MMRE, MdMRE and PRED(0.25). Experiments show that the DTF model outperforms the DT, as well as the MLR models in all evaluation criteria.

The rest of this paper is organized as follows: Section II introduces a background of terms used in this paper. Section III presents related work, whereas Section IV proposes the decision tree, decision forest tree and multiple linear regression models. In Section V, the proposed models are evaluated. Section VI lists the threats to validity, while Section VII concludes the paper.

## II. BACKGROUND

This section defines the main terms used in this paper that include decision tree, decision tree forest and evaluation criteria.

### A. Decision Tree (DT)

A DT is a logical model represented by a binary tree that illustrates how a target variable (aka dependent variable in regression models) is predicted using a set of predictor variables. A DT is composed of nodes, where the topmost node is known as the root and represents all the rows in a dataset. Then, each node is split into two nodes (aka child nodes) using a splitting variable. This is a recursive partitioning. Nodes that do not have child nodes are called terminals or leaf nodes. Leaf nodes carry the values of the target variable. The main advantage of a DT model is that it can help non-technical people to see the big picture of a certain problem. However, the main disadvantage of a DT model is that each node is locally optimized as opposed to global optimization of the whole tree. Furthermore, DT models might suffer from the overfitting problem, as well as from providing good accuracy in comparison to other models. The first DT program was developed by Morgan and Sonquist in 1963 and was called Automatic Interaction Detection (AID) [20]. This was followed by the THAID program in 1973 [21]. In this paper, the DTREG program [22] has been used to generate the DT model.

## B. Decision Tree Forests(DTF)

The DTF algorithm was first proposed by Leo Breiman in 2001 [23]. A DTF model consists of a collection of decision trees that grow in parallel. The predictions of the trees are combined to make the overall prediction of the forest trees. A DTF model is similar to a Treeboost model [24] in the sense that both models use a large number of trees, but the main difference between these two models is that in a Treeboost model, the trees are grown in series such that the output of one tree is fed into the next tree. In contrast, a DTF model is a collection of independent trees that are grown in parallel.

## C. Evaluation Criteria

To gauge the accuracy of the proposed models, we have used common evaluation criteria used in software estimation.

1. *MMRE:* This is a common criterion used to evaluate software effort estimation models [25]. The Magnitude of Relative Error (MRE) for each observation $i$ can be obtained as:

$$MRE_i = \frac{|Actual\ Effort_i - Predicted\ Effort_i|}{Actual\ Effort_i} \quad (1)$$

$$MMRE = \frac{1}{N}\sum_{i=1}^{N} MRE_i \quad (2)$$

MMRE is one of the most common methods used for evaluating prediction models; however, this method has been criticized by others such as [26], [11] and [27]. For this reason, we used a statistical significant test to compare between the median of two samples based on the absolute residuals. Since the absolute residuals were not normally distributed, the non-parametric statistical test Mann-Whitney U has been used to measure the statistical significance between different prediction models.

2. *MdMRE*: One of the limitations of the MMRE criterion is that it is sensitive to outliers. MdMRE has been used as another criterion because it is less sensitive to outliers.

$$MdMRE = median(MRE_i) \quad (3)$$

3. *PRED(x)*: The prediction level (PRED) has been used as a complimentary criterion to MMRE. PRED calculates the ratio of a project's MMRE that falls into the selected range (x) out of the total projects.

$$PRED(x) = \frac{k}{N}. \quad (4)$$

where $k$ is the number of projects where $MRE_i \leq x$ and $N$ is the total number of observations. In this work, PRED(0.25) has been used.

## III. RELATED WORK

This section presents associated work that is related to models that are based on decision tree that are used in software estimation.

Decision trees and fuzzy decision tree algorithms such as [16], [28], [29], [30] and [31] have been used in software effort estimation models. The authors in [28] developed two machine learning models; the regression tree and backpropagation neural network models. The COCOMO dataset was used in training whereas Kemerer's dataset was used in testing. The authors demonstrated that the proposed two models were competitive in comparison to other well-known models such as COCOMO, SLIM and Function Points. In [29], the authors address the issue of software cost estimation through fuzzy decision trees. The algorithms CHAID and CART were applied on the ISBSG dataset with fuzzy decision trees instances being generated and evaluated based on prediction accuracy. Huang et al. [30] proposed a fuzzy decision tree model for embedding risk assessment information into a software cost estimation model. The proposed approach was evaluated using the COCOMO dataset and yielded to better estimation results than the COCOMO model. The authors in [31] used machine learning techniques such as backpropagation neural networks, regression trees, radial basis functions and support vector regression methods for software effort estimation. The datasets used were USC, NASA and a Turkish dataset. The authors concluded that parametric models are insufficient for software cost estimation and the problem must be handled using an evolving system rather than a static one.

Elish [32] compares a Treeboost model (Stochastic Gradient Boosting model) with other neural and regression models using a NASA dataset that contains 18 projects. Nassif et al. [33] developed a Treeboost model for software effort estimation based on the Use Case Point model.

To the best of our knowledge, DTF models have not been used yet for software effort estimation. In this research, not only has a DTF model been developed to predict software effort, but a comparison between the DTF model against a DT model and a MLR has been conducted as well.

## IV. REGRESSION, DT AND DTF MODELS

This section introduces the MLR, the DT and the DTF models. We have used two published industrial data sets; namely, ISBSG and Desharnais for evaluation.

### A. Datasets Descriptions

1. **ISBSG**: In this work, ISBSG release 10 has been used. This release contains more than 4000 projects collected from many companies worldwide. Each project has several numerical and categorical attributes (features). Each project is rated as "A", "B" and "C" based on its quality, where "A" indicates the highest quality. A subset of 505 of "A" rating quality with nine numerical features has been selected in addition to the target variable 'Effort' as suggested in [12]. These features include 'AFP',

'input_count', 'output_count', 'enquiry_count', 'file_count', 'interface_count', 'add_count', 'delete_count' and 'changed_count.

2. **Desharnais**: The Desharnais dataset [34] is composed of a total of 81 projects developed by a Canadian software house in 1989. Each project has twelve attributes including 'project_number', 'team_experience', 'manager_experience', 'Year_End', 'Length', 'Effort', 'Transactions', 'Entities', 'PointsNonAdjust', 'Envergure', 'PointsAdjust' and 'Language'. All the attributes are numerical except 'Language' which is categorical. The attribute 'Language' has three values including '1' which corresponds to Basic Cobol, '2' which corresponds to Advanced Cobol and '3' which corresponds to 4GL. The attributes 'Length' and 'Effort' are both considered as target variables (dependent), thus the attribute 'Length' has been dropped. Similarly, the attributes 'project_number' and 'Year_End' were left out because they do not contribute to the target variable 'Effort'. The attributes 'PointsNonAdjust' and 'PointsAdjust' both correspond to software size so 'PointsNonAdjust' was dropped to eliminate the multicollinearity issue [35]. Out of the 81 projects, four projects contain missing attributes; therefore, only 77 complete projects are used. Descriptive statistics for the target variable 'Effort' of the ISBSG and the Desharnais datasets are depicted in Table I.

TABLE I.  DATASETS CHARACTERISTICS

| Characteristics | ISBSG | Desharnais |
|---|---|---|
| Min Effort (PH) | 668 | 546 |
| Max Effort (PH) | 14938 | 23940 |
| Mean Effort | 2828 | 4834 |
| Standard Deviation (Effort) | 2607 | 4188 |
| Skewness (Effort) | 2.09 | 2.04 |
| Kurtosis (Effort) | 4.51 | 5.30 |

*B. Multiple Linear Regression (MLR) Model*

The multiple linear regression model for each dataset was constructed using the 10-fold cross-validation technique. For illustration purposes, the models in Equations (5) and (6) as well as Figure 1 were developed based on the whole dataset.

1. MLR of ISBSG dataset:

Before a MLR was developed, a logarithmic transformation was applied on *Effort* and *AFP* (size) because *Effort* and *AFP* were not normally distributed. This method was recommended in [36]. Then, a stepwise regression was applied to see which attributes (independent variables) were statistically important at the 95% confidence level. The stepwise regression suggested that only three attributes are significant which are *AFP*, *enquiry* and *changed* as they contribute to the dependent variable *Effort*. The MLR equation of the ISBSG dataset is depicted in Equation 5.

$$\ln(Effort) = 5.94 + 0.31 \times \ln(AFP) \\ + 0.001 \times enquiry - 0.0005 \times changed. \quad (5)$$

The coefficient of determination $R^2$ is 21%. This means that only 21% of the variation in *Effort* can be explained by the independent variables *AFP*, *enquiry* and *changed*. This indicates that the MLR is not a good fit for the ISBSG dataset. The complete evaluation of the models is presented in Section V.

2. MLR of Desharnais dataset:

Regarding the MLR of the Desharnais dataset, a similar approach was applied as above with one main difference. The *Language* attribute in the Desharnais is categorical, so before we applied a MLR, this attribute was transformed into two dummy variables *L1* and *L2* such as the values of L1 and L2 are 1 and 0 respectively for Language 1 (Basic Cobol), 0 and 1 respectively for Language 2 (Advanced Cobol), 0 and 0 for Language 3 (4GL). The stepwise regression identified the attributes *PointsAdjust* and *Language* (which is represented by the dummy variables L1 and L2) as significant. Equation 6 shows the MLR equation of the Desharnais dataset.

$$\ln(Effort) = 1.69 + 0.97 \times \ln(PointsAdjusted) \\ + 1.36 \times L1 + 1.34 \times L2. \quad (6)$$

The coefficient of determination $R^2$ is 77.7%. This means that 77.7% of the variation in *Effort* can be explained by the independent variables *PointsAdjusted*, *L1* and *L2*.

*C. Decision Tree Model (DT)*

Likewise, the DT model was developed from the ISBSG and the Desharnais datasets using the 10-fold cross-validation technique.

1- Decision Tree of ISBSG dataset:

The parameters of the DT model were chosen so that the error is minimal with one exception which is tree pruning. As explained in Section VI, pruning will simplify the model but will deteriorate the accuracy. The parameters of the DT model are as follows:
Maximum splitting levels: 10; Type of analysis: Regression; Splitting algorithm: Least squares; Variable weights: Equal; Minimum size node to split: 10; Minimum rows allowed in a node: 5; Tree pruning and validation method: Cross validation; Number of cross-validation folds: 10. Figure (1) shows the DT model of the ISBSG dataset. The tree was pruned based on the minimum value of the cross-validation error. Please note that the decision tree algorithm suggested that only the attributes *AFP* and *enquiry* were significant.

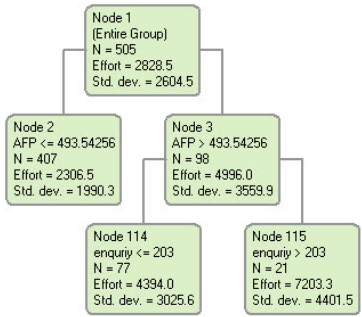

Figure 1. Decision Tree ISBSG

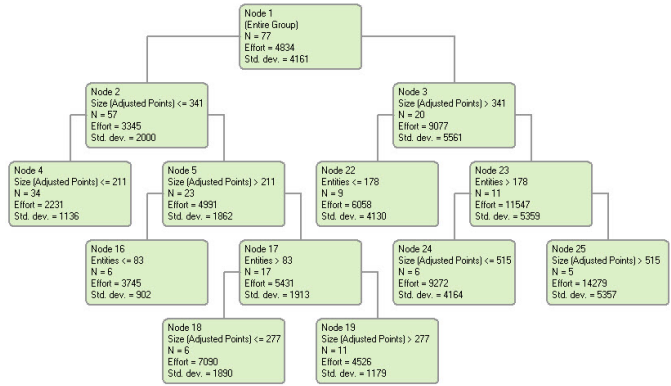

Figure 2. Decision Tree Desharnais

2- Decision Tree of Desharnais dataset:

Likewise, the DT model of the Desharnais dataset (Figure 2) was developed based on the same parameters described above. The splitting variables (most significant variables) are *PointsAdjusted* and *Entities*, and this contradicts the significant variables (*PointsAdjusted* and *Language*) that were identified by the stepwise regression.

This shows that significant variables need not be the same in all models, specifically that the MLR is a linear model whereas the DT model is a non-linear one.

### D. Decision Tree Forest (DTF) Model

The DTF model was developed from the ISBSG and the Desharnais datasets using the 10-fold cross-validation technique.

1- DTF of ISBSG dataset:

The parameters of the DTF model are as follows:
Maximum trees in Decision Tree Forest: 500; Maximum splitting levels: 100; Type of analysis: Regression; Variable weights: Equal; Minimum size node to split: 2; Tree validation method: Out of Bag (OOB).

The DTF model uses the "out of bag" data rows for validation. This provides an independent test without requiring a separate data set or holding back rows from the tree construction. The number of predictors used for each split was 2 out of 8 (square root of the number of predictors) as recommended by Breiman [23]. The full forest has 500 trees and the Maximum depth of any tree in the forest was 29.

2- DTF of the Desharnais dataset:

The DTF model of the Desharnais dataset was developed using the same parameters as the ISBSG dataset. This has yielded for the forest to be 500 trees and the maximum depth of any tree in the forest to be 17.

## V. MODEL EVALUATION AND DISCUSSION

This section presents the evaluation of the MLR, DT and DTF models in the training stage based on the MMRE, MdMRE and PRED(0.25) criteria as well as the median of absolute residuals as shown in Table II.

TABLE II. MODEL EVALUATION

| Criteria | MLR | DT | DTF |
|---|---|---|---|
| MMRE (ISBSG) | 0.58 | 0.49 | 0.17 |
| MdMRE (ISBSG) | 0.63 | 0.56 | 0.18 |
| PRED(0.25) ISBSG | 7.8 | 23 | 83 |
| Median of Absolute Residuals | 856 | 1148 | 339 |
| MMRE (Desharnais) | 0.32 | 0.56 | 0.25 |
| MdMRE (Desharnais) | 0.27 | 0.28 | 0.14 |
| PRED(0.25) Desharnais | 46 | 44 | 72 |
| Median of Absolute Residuals | 871 | 1044 | 563 |

### A. Discussion

Table II shows that the DTF model outperforms the MLR as well as the DT models, based on all criteria. As a comparison between the DT and the MLR models, we see that the DT model surpasses (lower MMRE, MdMRE values and higher PRED values) the MLR when the ISBSG dataset was used. In contrast, the MLR outperforms the DT model when the Desharnais dataset was used. To confirm the robustness of the DTF model, we measured the non-parametric Mann-Whitney U test between the DTF model and the other two models based on absolute residuals as shown in Table III. Results show that the DTF model is statistically significant at the 95% confidence level (p value < 0.05).

TABLE III. MANN-WHITNEY U TEST

| Models | Mann-Whitney (p-value) |
|---|---|
| DTF vs MLR (ISBSG) | 0.00 |
| DTF vs DT (ISBSG) | 0.00 |
| DT vs MLR (ISBSG) | 0.00 |
| DTF vs MLR (Desharnais) | 0.01 |
| DTF vs DT (Desharnais) | 0.00 |
| DT vs MLR (Desharnais) | 0.48 |

## VI. THREATS TO VALIDITY

1- The DT model was pruned to reduce the complexity of the model so that it can be easily comprehended. However, this would decrease the accuracy of the model. The precision of the DT model would have been improved if all the variables were used (non-pruned model).

2- Despite the fact that the DTF model gives good results, this model and all decision tree based models cannot be used to predict an effort of a project which is beyond the training dataset. For instance, Node 115 in Figure1 states that the effort of any project who's AFP (size) is greater than 493 and whose enquiry is greater than 203 is 7203.3 person-hours. Thus, irrespective of the splitting variables values, the maximum estimated effort of this DT model will be 7203.3 person-hours.

VII. CONCLUSIONS

This paper compared a decision tree forest (DTF) model with a decision tree (DT) model, as well as a multiple linear regression (MLR) model for software effort prediction. The three models were developed using 10-fold cross-validation technique using the ISBSG and the Desharnais datasets. The evaluation criteria used were MMRE, MdMRE and PRED(0.25). Results show that the DTF model outperforms the other two models based on all evaluation criteria. The robustness of the DTF model was confirmed using the non-parametric Mann-Whitney U Test. Based on these results; we conclude that the DTF model can be used with promising results to predict software effort.